\documentclass[11pt,letterpaper]{article}
\usepackage{emnlp2017}
\usepackage{times}
\usepackage{latexsym}
\usepackage{algorithm}
\usepackage[noend]{algpseudocode}
\usepackage{url}

\usepackage{geometry, caption, subfig,amsmath, enumitem, setspace,graphics,hyperref, comment}
\usepackage{graphicx}
\usepackage{rotating}
\usepackage{amsthm}
\theoremstyle{definition}
\newtheorem{definition}{Definition}[section]

\usepackage{color}
\usepackage{xcolor}
\definecolor{orange}{RGB}{255,127,0}

\emnlpfinalcopy

\def\emnlppaperid{\#\#}

\title{A Computational Framework for Multi-Modal Social Action Identification}

\author{L. Jason Anastasopoulos\thanks{Princeton University; University of Georgia, \href{ljanastas@princeton.edu}{ljanastas@princeton.edu}}  \and Jake Ryland Williams\thanks{Drexel University, Department of Information Science, College of Computing and Informatics}}

\date{\today}

\begin{document}

\maketitle

\begin{abstract}
  \fontsize{10}{11}\selectfont
  We create a computational framework for understanding social action and
  demonstrate how this framework can be used to build an open-source event detection tool
  with scalable statistical machine learning algorithms and
  a subsampled database of over 600 million geo-tagged Tweets
  from around the world.
  These Tweets were collected between April 1st, 2014 and April 30th, 2015,
  most notably when the Black Lives Matter movement began.
  We demonstrate how these methods
  can be used diagnostically---by researchers,
  government officials and the public---to
  understand peaceful and violent collective action at
  very fine-grained levels of time and geography.  
\end{abstract}

As violent forms of collective action continue to erupt around the globe, there is a growing need to understand the conditions leading to them.
Violent collective action not only poses significant dangers to the
individuals and communities directly affected by it,
but also tends to diminish the legitimacy of the causes associated with it~\cite{HuetVaughn2013a,Porta1986a}, potentially hindering needed social change.
For researchers, an automated means of identifying violent and peaceful collective action
can provide a rich source of data that
will expand the scope of knowledge and understanding of modern social movements and
overcome the inherent data availability limitations which have
restricted the study of social movements to
either detailed case studies~\cite{Tilly2008a,Tilly2015a} or
news media sources~\cite{Earl2004a}.
From a public safety perspective,
citizens can benefit from a tool which could warn them about
areas in which violent activity is currently occurring or
where it is likely to occur so that these areas can be avoided.
From a mobilization perspective,
utility may also be derived from a tool which can
help potential participants
discriminate between the types
of activities that they would like to participate in.

In this paper, we create a computational framework for identifying
different forms of social action and use this framework as the
basis for a series of scalable event-detection algorithms that
can be used to identify, track and study violent and non-violent collective action at
fine-grained temporal and geographic levels.
We build these algorithms using a subsample of an over 600 million geo-coded Tweet database
collected between April, 1st 2014 and April 30th, 2015.
Using these algorithms, we explore the
linguistic and spatial features of Tweets,
describe those related to peaceful and forceful forms of collective action,
and then demonstrate how these algorithms can be used to build databases which
contain metrics of collective action. 

\section*{Framework for Social Action Identification}
Building a computational framework for social action identification requires identification of actions or observations
conducted by individuals on behalf of
or in concert with collectives.
On social media, individuals post content (text, images, etc.)
with the explicit knowledge and intention
that others will be engaging with the information in some way.
Thus, all activity on social media might be considered
as a Weberian social action~\cite{Weber1922a}.
This generality initiates our formalization:
\theoremstyle{definition}
\begin{definition}{Social actions,}
  $\mathcal{S}$,
  form the subset of all actions in which individuals participate,
  $\mathcal{A}$,
  where \emph{consideration is made for the potential actions and reactions of others}.
\end{definition}
\noindent Thus, actions outside of $\mathcal{S}$---asocial actions---fall
entirely outside of our study, given our view through social media.

While all acts performed on social media
fall in $\mathcal{S}$ by the Weberian definition,
much of the content produced on these platforms
does not comport with notions of
social movements and collective action.
These are our focus, so how are they distinguished?
To focus in on social movements,
we follow Tilly's notion~\cite{Tilly1984a}:
\begin{quote}
  \textit{...a sustained series of interactions between powerholders and persons successfully claiming to speak on behalf of a constituency lacking formal representation, in the course of which those persons make publicly visible demands for changes in the distribution or exercise of power, and lack those demands with public demonstrations of support.}
\end{quote}
Distilling the inherently \emph{political}
nature of these types of actions,
which hinge on representation, constituency, and power dynamics,
we refine $\mathcal{S}$ as follows:

\begin{figure*}[ht]
  \centering
  \includegraphics[width=\textwidth]{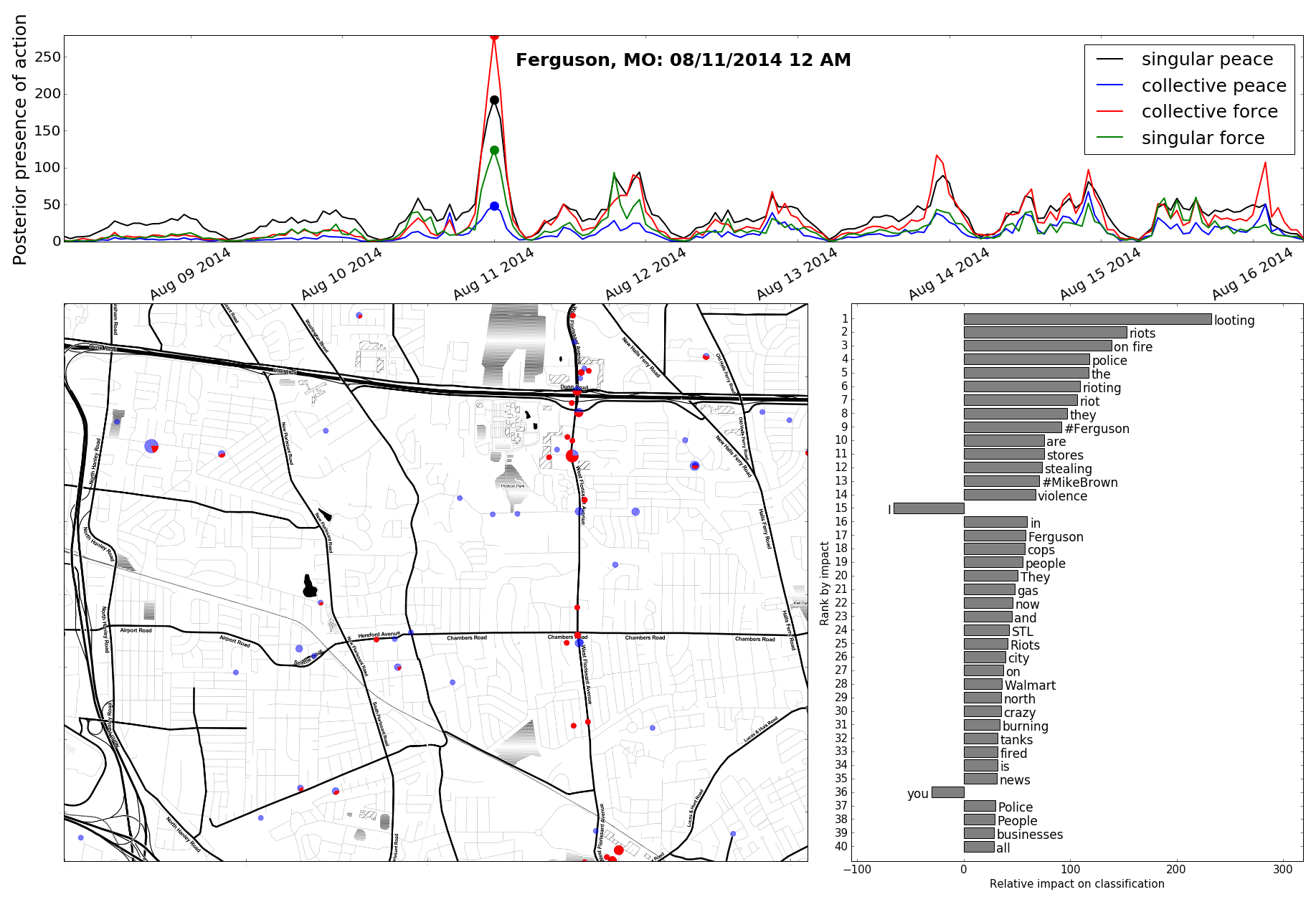}
  \caption{
    Above. Time series showing the total presence of
    social action types in Ferguson, MO
    the week after the shooting of Michael Brown
    by Officer Darren Wilson on August 9, 2014.
    The presence of each action type is determined by our
    adept Bayes classifier and
    measured as the sum of posterior probabilities
    over all tweets from each hour in the
    plotted span of time.
    Left. Map of Ferguson, MO
    depicting clusters of collective force activity
    over one hour around 12 AM, on August 11th.
    The size of each cluster-circle
    represents the
    area from which tweets emerged
    (not the number of tweets contained),
    and the portion of each circle colored red
    indicates the portion of tweets classified to
    represent the collective force action.
    Right. A phrase shift showing the
    most impactful features present in
    all tweets classified as being representative of collective force.
    Phrases on the right pull the classifier toward a positive classification,
    and phrases on the left pull the classifier towards a negative classification.    
  }
  \label{fig:Ferguson}
\end{figure*}

\theoremstyle{definition}
\begin{definition}{Political actions,}
  $\mathcal{P}\subseteq\mathcal{S}$,
  are the subset of social actions
  in which individuals and groups
  exercise or express power
  on behalf of themselves or others.
\end{definition}

Tilly's formalization around power, constituency, and representation
winnows the actions represented on social media to, perhaps,
a more interesting subset.
Furthermore, the distillation of Tilly's notions
lead us to explore two natural refinements
that are hinted at in his definition.
These emerge from two questions:
\begin{itemize}
\item[$1)$] What are the scales of acting entities?
\item[$2)$] In what manners do entities exercise power?
\end{itemize}
We see specific political actions as
falling along one-dimensional spectra
with respect to each question; e.g.;
(1) small-to-large scales, ranging from individuals, to teams, to collectives, to states;
or;
(2) civil-to-unruly manners, ranging from exhibition, to negotiation, to declaration, to enforcement.
A specific numeric value on each spectrum
is difficult to identify,
so we approach these initially
by establishing criteria that
represent each as a simple dichotomy:
\theoremstyle{definition}
\begin{definition}{Actor scales}
  refine the space of political actions
  into \emph{singular} actions, $\mathcal{I}$,
  conducted by individuals,
  and \emph{collective} actions, $\mathcal{C}$,
  conducted by groups operating in unison.
  Singular and collective actions
  disjointly partition the political actions:
  $\mathcal{C}\cup\mathcal{I} = \mathcal{P}$;
  $\mathcal{C}\cap\mathcal{I} = \emptyset$.
\end{definition}
\theoremstyle{definition}
\begin{definition}{Action manners}
  refine the space of political actions
  into \emph{peaceful} actions, $\mathcal{U}$,
  for which all powers exercised
  respect the wills of all engaged parties
  and \emph{forceful} actions, $\mathcal{V}$,
  for which some power is exercised
  in violation of the will of another.  
  Likewise, peaceful and forceful actions
  disjointly partition the political actions:
  $\mathcal{U}\cup\mathcal{V} = \mathcal{P}$;
  $\mathcal{U}\cap\mathcal{V} = \emptyset$.
\end{definition}

While each of our refinements of the political actions
partition $\mathcal{P}$,
scales and manners interact in non-trivial ways.
For example, groups may act forcefully,
and individuals may act peacefully, etcetera.
Thus, the full range of actions
distinguished under our framework
is now a partition of $\mathcal{P}$ into four \emph{modes}:
\begin{definition}{Singular peace}
  actions, $\mathcal{I}\cap\mathcal{U}$,
  are those conducted by individuals
  that respect the wills of others,
  including negotiations, arguments,
  condemnation, and
  expressions of empathy and support.
\end{definition}

\begin{figure*}[ht]
  \centering
  \includegraphics[width=\textwidth]{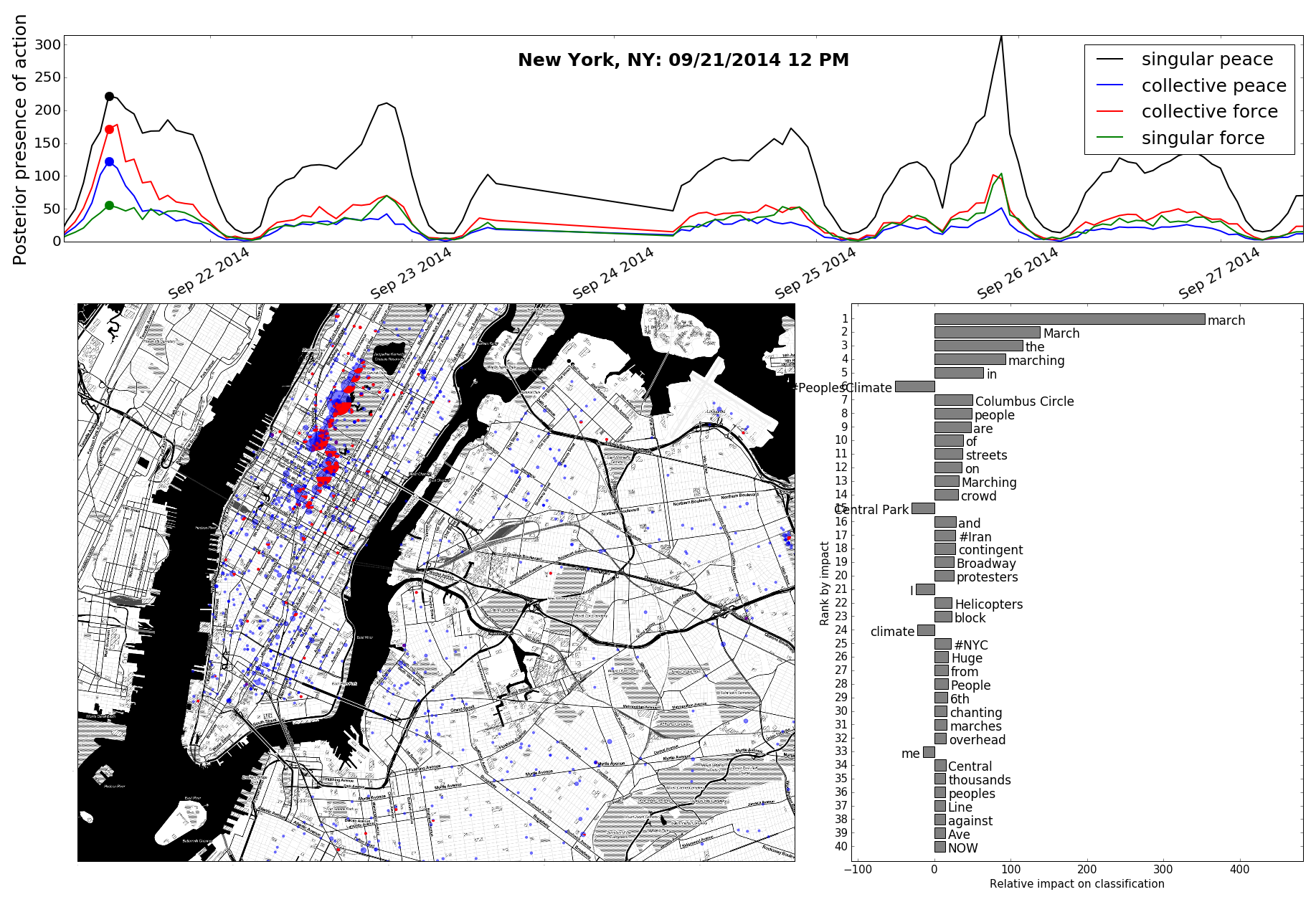}
  \caption{
    Above. A time series showing the total
    presence of social action types during the week of the
    People’s Climate March which began on September 21, 2014.
    Left.  Map of New York, NY depicting
    clusters of collective force and collective peace activity
    over one hour around 12 PM on September 21st
    during a climate change protest.
    The size of each cluster-circle
    represents the area from which tweets emerged
    (not the number of tweets contained),
    and the portion of each circle colored red
    indicates the portion of tweets classified to
    represent the collective force action.
    Right. A phrase shift showing the most impactful features
    present in all tweets classified as being representative of collective force.
    Phrases on the right pull the classifier toward a positive classification,
    and phrases on the left pull the classifier towards a negative classification.    
  }
  \label{fig:NewYork}
\end{figure*}

\begin{definition}{Singular force}
  actions, $\mathcal{I}\cap\mathcal{V}$,
  are those conducted by individuals
  that violate the wills of others,
  including assassinations, slander,
  shootings and other individually conducted assertations.
\end{definition}
\begin{definition}{Collective peace}
  actions, $\mathcal{C}\cap\mathcal{U}$,
  are those conducted by groups
  that respect the wills of others,
  including vigils/singing, lawful congregation,
  food/blood drives, and petitions.
\end{definition}
\begin{definition}{Collective force}
  actions, $\mathcal{C}\cap\mathcal{V}$,
  are those conducted by groups
  that violate the wills of others,
  including suppression, blockades,
  unlawful congregation, and rioting.
\end{definition}
\noindent Note that some actions may be categorized differently,
according to the prevailing circumstances
in which they are carried out, e.g.,
the action of
a vandal acting alone would
fall under singular force,
while another's performance in a riot
would fall under collective force.

\begin{figure*}[ht!]
  \centering
  \includegraphics[width=\textwidth]{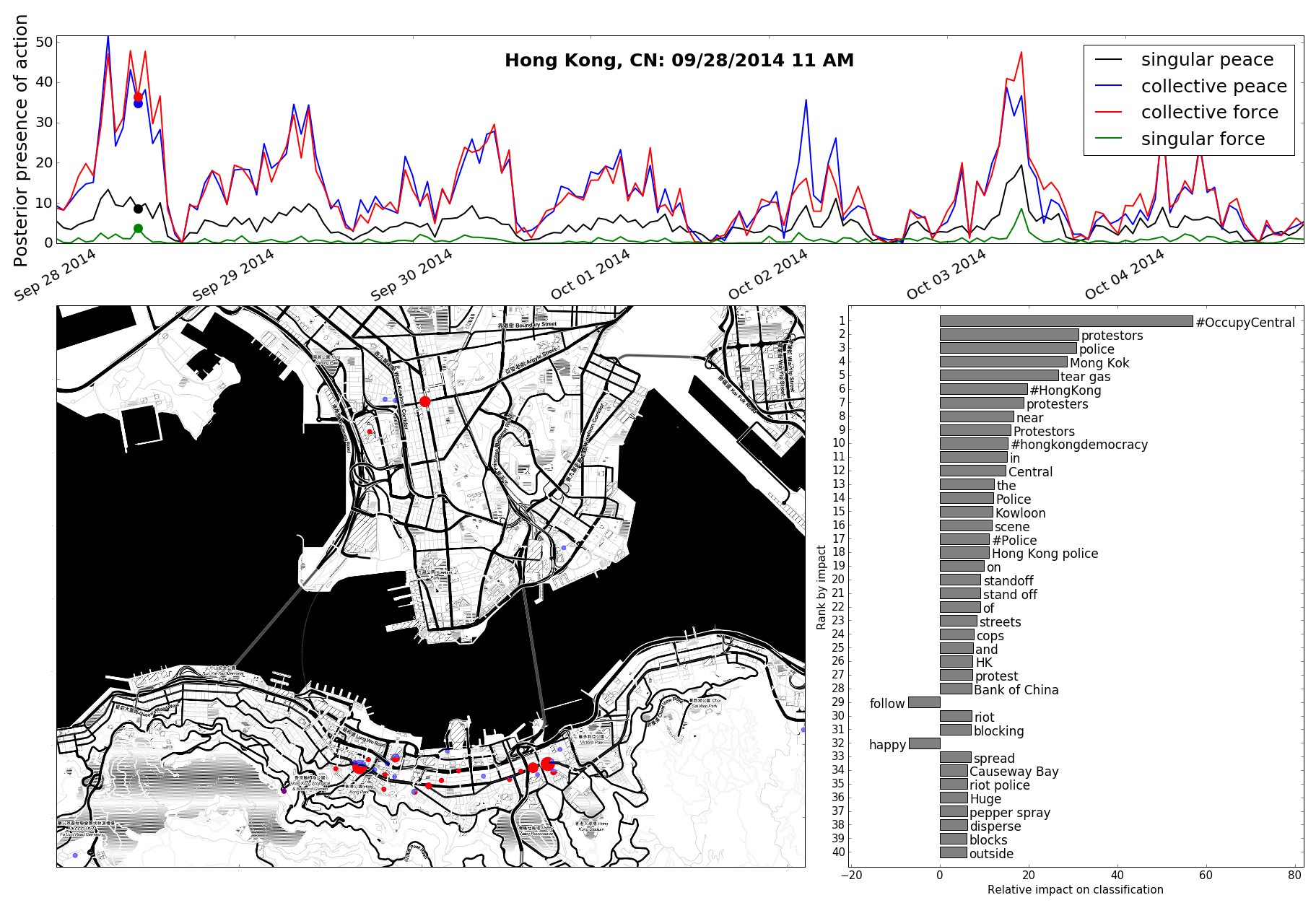}
  \caption{
    Above. A time series showing the total presence of social action types
    during the week of the Occupy Central with Love and Peace movement
    which began on September 28th, 2014 Left.
    Map of Hong Kong depicting clusters of collective force activity
    over one hour around 11 AM, on September 29th.
    Right. A phrase shift showing the most impactful features
    present in all tweets classified as being representative of collective force.
    Note: points and bars represent analogous quantities to those
    in Figs.~\ref{fig:Ferguson}~and~\ref{fig:NewYork}.
  }
  \label{fig:HongKong}
\end{figure*}

\section*{Building a social action classifier}

To explore the measurement of social action under our model,
we utilize a database of over $600$ million
high-precision geographically-tagged messages from
the Twitter social network collected over April 2014–April 2015,
which most notably covers the
beginning of the Black Lives Matter (BLM) movement extensively.
This dataset was collected from Twitter’s public (spritzer) API
and is of particular importance for Twitter’s policies
around location tagging at the time of collection.

At the time, when a user opted in for location tagging from a mobile device,
a tweet sent would automatically be
accompanied by high-precision latitude and longitude coordinates.
Since then, Twitter enacted a policy that
resulted in the adjustment of their system to
default the option of soft-locations,
which users specify
(for example,
one could set their location to Philadelphia
and then go on to tweet from anywhere
else in the world,
with tweet meta-data always listed as Philadelphia).
Since the rate at which tweets were geo-tagged was
approximately 1\% at the time of collection,
and the (1\% stream) public API was
restricted to only location-tagged tweets,
it is arguable that this dataset
constitutes a near-complete collection of geo-tagged tweets during this period.

On account of the vastness of the Twitter database,
we construct a filter that
enables us to identify sets of Tweets
more likely related to protest activity over the time period.
While this kind of information can be
technically accessed through newspaper accounts of protest activity,
identifying the locations and exact times that protests took place
around the world using newspaper accounts
would require collecting a massive database of newspaper articles
in different languages from around the world.

Instead, we leveraged an \textit{Associated Press} image database
containing thousands of images and
extracted the relevant metadata contained therein,
which included the
exact time and location that photos of protest activity were taken.
With this information,
we build a Tweet ``protest filter'' for
our coding of social action
(assuming this sample to be particularly potent
in representation of the actions of interest)~\cite{AP2016}.

Using approximately 10\% (18,000) of the tweets
sampled from the AP-filtered protest times and locations,
in addition to all tweets from Alameda county, California
on the night of Nov. 24th
(a place and time that was known internationally to have violent protest activity),
we coded tweets individually
for the presence of the four modes of social action.
In particular, we accept that tweets may
represent any number of the four types of action.
From the total 22,626,
a breakdown of the positively coded tweets is given in Tab.~\ref{tab:evaluation}.

We use the coded tweets as input for binary na\"{i}ve Bayes classifiers,
which, for each of the four action modes, run in parallel.
These standard na\"{i}ve Bayes classifiers
are also modified with
an enhanced input feature space of
both single and multiword expressions.
This process is accomplished 
through a recently developed~\cite{Williams2017a}
multiword expression segmentation method,
which, intuitively, bases our classifieres
on integrated collections of words and phrases.
We refer to the resulting systems
as ``adept'' Bayes classifiers,
whose features have two notable advantages:
\begin{itemize}
\item[1.] independent, semantic accuracy, and
\item[2.] out-of-context human interpretability. 
\end{itemize}
So for example,
basing the adept classifier on
the expression \emph{tear gas}
(see $5^\text{th}$ ranked word in
Fig.~\ref{fig:HongKong}, right)
sidesteps confounding
statistical effects in the frequencies
of the words \emph{tear} and \emph{gas},
and at the same time may be interpreted by
a diagnostician to appropriately mean
a crowd suppression device.
Interpreted separately,
these words might indicate an epidemic of indigestion.

\begin{figure*}[ht!]
  \centering
  \includegraphics[width=\textwidth]{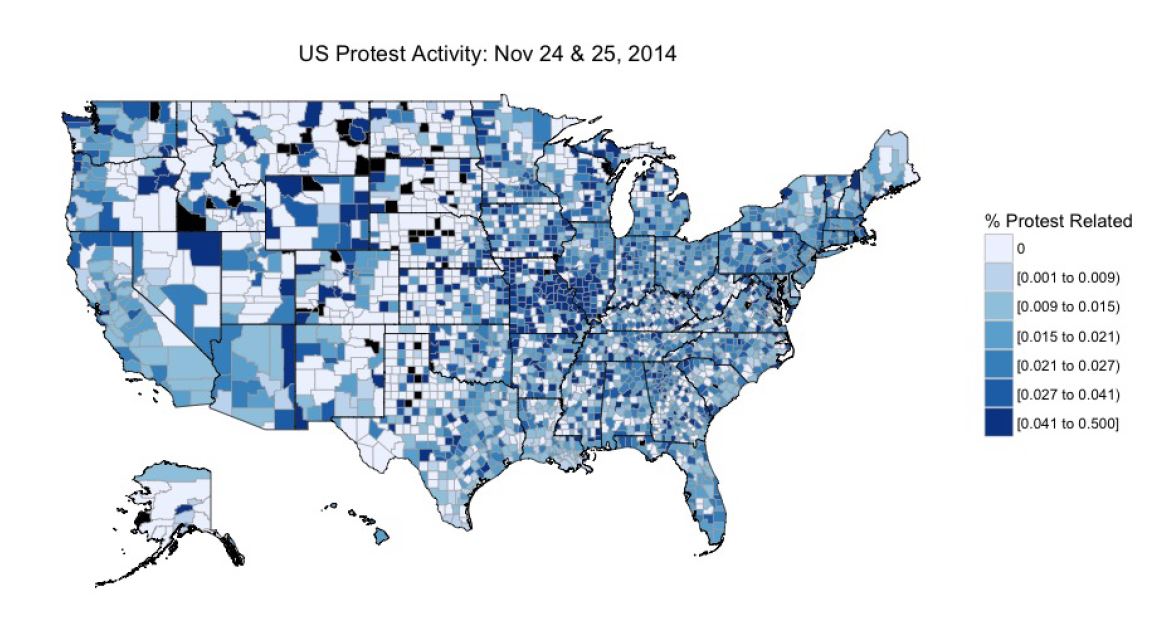}
  \caption{
    Ferguson-related political action
    across US counties on November 24th and 25th, 2014
    as measured by the percentage of Tweets
    related to any form of political action.
  }
  \label{fig:politicalActivity}
\end{figure*}

\section*{From classifier to diagnostic utility}

In addition to improving the Bayes classifier used in our experiments,
the usage of phrases as features
allows for greater interpretability of classifications.
Our adept Bayes classifier has an
advantage of being explorable,
as a ``white box'' method that can be opened to
show the features most relevant to classifications.
In particular,
looking at a document as  a bag of phrases
$d = {w_1, w_2, \cdots , w_N}$,
counted with  frequencies
${f(w_1), f(w_2), \cdots , f(w_N )}$,
their impact on the adept Bayes classification
is largely due to the likelihood function,
$\Lambda$ (determined in training),
which, often computed as a sum of logarithms,
is linear in frequencies:
\begin{equation}
  \label{eq:likelihood}
  -\sum_{i = 1}^Nf(w_i)\log_{10} \Lambda(w_i\mid c).
\end{equation}
Note: $c$ is the mode's presence (positive/negative),
and the terms are negated for an entropic framing.
If a diagnostician wishes to understand why
a tweet was classified as
positive ($c_{+}$) over negative ($c_{-}$),
the difference may be computed:
\begin{equation}
  \label{eq:difference}
  -\sum_{i = 1}^Nf(w_i)\left(\log_{10} \Lambda(w_i\mid c_{+}) - \log_{10} \Lambda(w_i\mid c_{-})\right).
\end{equation}
Such a difference affords a ranking of
features by the (absolute) terms of the sum,
i.e., each word, $w_i: i=1,\cdots, N$,
can be compared according to the relative impact on classification:
\begin{equation}
  \label{eq:ranking}
  f(w_i)\left(\log_{10} \Lambda(w_i\mid c_{+}) - \log_{10} \Lambda(w_i\mid c_{-})\right).
\end{equation}
Thus, for diagnostic value we display the
ranked values of Eq.~\ref{eq:ranking} along a
vertical bar plot, which we call a phrase shift
(see Figs.~\ref{fig:Ferguson},~\ref{fig:HongKong},~and~\ref{fig:NewYork}).

\begin{table*}[t]
  \centering
  \caption{
    Tenfold cross-validation results
    from application of the na\"{i}ve Bayes classifier
    for the different modes of action.
    These results are also presented for less-refined modes.    
    Each labeled row indicates the number of positive codings
    in the training data set (\textbf{Abundance}),
    and the $F_1$-optimal posterior probability (\textbf{Threshold})
    for classification,
    in addition to its the corresponding values of
    precision (\textbf{P}), recall (\textbf{R}) , and combined $F_1$.
    Out-of-domain evaluations are presented parenthetically,
    adjacent to their corresponding in-domain values.
  }
  \label{tab:evaluation}
  \begin{tabular}{l|lllll}
    \bf Action & \bf Abundance & \bf Threshold & \bf P & \bf R & \bf $F_1$ \\
    \hline
    \bf Collective force & 795 (99) & 0.08 (0.35) & 74.08 (64.44) & 76.01 (58.59) & 74.94 (61.38)\\
    \bf Collective peace & 474 (111) & 0.78 (0.04) & 51.92 (45.87) & 55.25 (45.05) & 53.29 (45.45)\\
    \bf Singular force & 351 (6) & 0.92 (0.42) & 57.39 (0.25) & 41.19 (16.67) & 47.38 (20)\\
    \bf Singular peace & 1,823 (96) & 0.85 (0.11) & 73.52 (44.68) & 67.61 (43.75) & 70.38 (44.21)\\
    \bf Collective & 1,116 (168) & 0.79 (0.17) & 74.54 (75.91) & 68.8 (61.90) & 71.48 (68.20)\\
    \bf Singular & 1,951 (101) & 0.87 (0.36) & 71.44 (41.18) & 68.57 (55.45) & 69.90 (47.26)\\
    \bf Force & 1,107 (103) & 0.71 (0.21) & 66.94 (60.19) & 67.54 (63.11) & 67.22 (61.61)\\
    \bf Peace & 2,092 (178) & 0.88 (0.23) & 71.5 (53.69) & 72.2 (61.24) & 71.78 (57.22)\\
    \bf All & 2,596 (226) & 0.93 (0.24) & 80.71 (65.52) & 74.42 (75.66) & 77.39 (70.23)\\
  \end{tabular}
\end{table*}

\section*{Evaluation}

We examine the performance of our classifier by
performing a tenfold cross-validation on the coded tweets data set.
The results of this validation are recorded in Tab.~\ref{tab:evaluation}.
Treating the Bayes posterior probability as a tunable threshold for classification,
we measure precision and recall,
and optimize the threshold probability over $F_1$
to tune each given classifier.
Observing these results,
we see that collective force is, individually,
the best predicted action type.
This is encouraging, as collective force often represents the most serious actions.
While classifier performance at
predicting collective peace and singular force is lower,
we do see that the most prevalent type of action, singular peace, is predicted well.
When the classifiers are collapsed to less-specific types of action
(Collective, Singular, Peace, and Force)
performance  decreases from the best cases
(singular peace and collective force),
but when all action types are combined (All),
we see a significant performance improvement in all measures.

While we cannot quantify our model’s performance in
application to real-time data yet,
we can hint at its performance on out-of-domain data
by separating known distinct events in the training data.
In addition to the BLM movement,
the coded data significantly
cover a portion of the Hong Kong democracy protests.
Using the data from Hong Kong for testing,
and all other data for training,
we see somewhat different results (see Tab.~\ref{tab:evaluation}, parentheticals),
and generalize to
note some potential challenges with
using this data
(that only covers a limited set of events)
to build a classifier
and apply it to data representing unknown and unforeseen events.

First, the subject matter (democracy) from the Hong Kong tweets
is very different from that of the BLM movement (institutionalized racism),
making the discourse present in the singular peace test tweets
largely unrelated to that from training.
As a result, this previously predictable category
now exhibits substantially decreased performance.
Furthermore, of the nearly 900 tweets coded from Hong Kong,
only 6 were found to be representative of singular force,
so there is essentially nothing to predict for this category.

For the collective actions we see very different numbers,
especially for collective force.
This is likely as a result of the
similar collective tactics
employed on both sides of both movements
(e.g., blockades, non-lethal pacification, etc.).
When the different action type are collapsed,
we see more and more performance improvements,
indicating that the collapsed categories
may be the most reliable.
However, since the Hong Kong tweets are
actually part of the training of the overall classifier,
we note that the performance of our model
when applied to real-time data
will likely be better than that reported in Tab.~\ref{tab:evaluation} (parentheticals),
and importantly,
for the most serious type of action---collective force---our
results in performance are largely upheld.

\section*{Interpretation}

To exhibit the manner in which our classifier might be used,
we apply our trained classifier to data from outside of training,
taken from Ferguson, MO. during the initial wave of protest activity, over August of 2014.
In Fig.~\ref{fig:Ferguson} we plot a time series of this period,
showing the abundance of the four types of social action,
as measured by the
sum of posterior probabilities of all tweets under the application of our classifiers.
Here, it can be seen that the
largest spikes occurred on the first night of protesting (Aug. $10\textsuperscript{th}$).
While singular peace (black line) exhibits a substantial,
periodic signature even under normal circumstances
(the discourse it covers is regular and common),
collective force (red line) emerges aberrantly during the protest events,
overshadowing the presence of the other action types.

Taking a closer look at the presence of collective force at the largest spike,
we zoom in to a map of the first night of protesting in Fig.~\ref{fig:Ferguson},
and plot clusters of tweets with the
positive classifications represented as proportional areas.
Here, we can see a larger cluster just south of the freeway,
on West Florissant Avenue,
which corresponds to the
time and location of the burning of the QuikTrip convenience store and gas station
(set to fire by protesters).
This action is actually hinted at in the phrase shift (bar plot, right),
by terms such as ``burning'' and ``on fire.''
While the first night of protesting was violent and unexpected,
the actions that took place were spread out,
and involved fewer mass confrontations with the police,
which later became more militarized and
can be observed in figures
depicting subsequent evenings.

We additionally present the result of our model’s application to
the Hong Kong democracy protests that
lasted for approximately two months in the fall of 2014.
On the map in Fig.~\ref{fig:HongKong}, we see clusters of
collective force activity at the three main protest sites:
Admiralty, Causeway Bay, and Mong Kok.
Tactics similar to those reported by Twitter users
during the Ferguson protests were
employed by the Hong Kong police as well,
as is indicated by a phrase shift (Fig.~\ref{fig:HongKong}, right)
that shows highly-impactive phrases such as
``tear gas,'' ``stand off,'' and ``riot  police.''
So for the collective force category,
we see a large degree of accord in the
lexical features that indicate the presence of the action
(which we quantify below, in Tab.~\ref{tab:evaluation} (parentheticals)),
indicating the possibility of
applicability to out-of-domain data, and future events.

\section*{Building a Protest Activity Database}

While the literature discusses many aspects of protests,
one of the central political questions surrounding protest activity
involves its effects on public opinion~\cite{Dunway2010a,Wallace2014a,Branton2015a} and
its ability to influence policymakers.
The effects of protest activity on public opinion are
especially important to understand because
the former typically influences the latter.
To construct measures of protest activity
researchers have mostly relied on compiling databases of
newspaper coverage of protest activity~\cite{Earl2004a}.
While these databases arguably capture
some of the most impactful protest activity,
they have limited the study of protests to those picked up by media sources.

Here, we demonstrate that our software can be used to
build a database of measurable protest activity
at the county level in the United States.
Using our software, we explore patterns of protest activity
across the United States shortly after
the grand jury acquittal of Officer Darren Wilson in Ferguson, MO on November 24th
and construct measures of protest activity using a subsample of 3.5 million classified Tweets.
The Fig.~\ref{fig:politicalActivity} map contains measures of
overall political activity
across the United States---notably centralized in Missouri---after
the acquittal of Darren Wilson on November 24th, 2014.

\section*{Discussion}
As collective action in the digital age increasingly becomes a phenomenon which occurs simultaneously on social media and in geographic spaces, a theory which is able to map textual data and metadata onto events occurring on the ground provides a means by which these data can be harnessed to better understand the evolution of modern social movements.
In this paper, we present a framework for identifying social actions which we argue accomplishes this task and demonstrate how this framework can be used to identify collective action and other social phenomena with high precision in a machine learning context. 
While the software and model discussed above were
constructed using Twitter data,
this method can be applied to any text-based form of
real-time streaming social media.
As such, this paper adds to a growing body of work
focused on the automatic detection of events
from social media streams in general,
such as~\cite{Sayyadi2009a,Aggarwal2012a,Nurwidyantoro2013a,Zhou2014a,Dong2015a,Wang2016a,Ramakrishnan2014a},
to name a few.
The information we have depicted in Figs.~\ref{fig:Ferguson},~\ref{fig:NewYork},~and~\ref{fig:HongKong}
serve as examples of the diagnostic utility that our developed framework and methods can provide.

\bibliography{Computation-Social-Action-EMNLP}
\bibliographystyle{emnlp_natbib}

\end{document}